\documentclass[letter]{IEEEtran}
\usepackage{cite}
\usepackage{graphicx}
\usepackage{times}
\providecommand{\keywords}[1]{\textbf{\textit{Index terms---}} #1}
\usepackage{latexsym}
\usepackage{bm}
\usepackage{amssymb,amsmath,mathtools}
\usepackage[center]{caption2}
\usepackage{stfloats}
\usepackage{cases}
\usepackage{url}
\usepackage{array}
\usepackage{setspace}
\usepackage{fancyhdr}
\usepackage{color}
\usepackage{subfigure}
\usepackage{chemarrow}
\usepackage{extarrows}
\usepackage{algorithm}
\usepackage{algorithmic}
\usepackage{enumerate}
\newtheorem{theorem}{\underline{Theorem}}
\newtheorem{lemma}{\underline{Lemma}}

\def\proof{\noindent\hspace{2em}{\itshape Proof: }}
\def\endproof{\hspace*{\fill}~$\square$\par\endtrivlist\unskip}
\allowdisplaybreaks[4]

\makeatother
\begin{document}
\title{{Intelligent Reflecting Surface Aided Multicasting with Random Passive Beamforming}
\thanks{Q. Tao and C. Zhong are with the College of Information Science and Electronic Engineering, Zhejiang University, China (email: \{taoqin,caijunzhong\}@zju.edu.cn). S. Zhang and R. Zhang are with the Department of Electrical and Computer Engineering, National University of Singapore (email: \{elezhsh,elezhang\}@nus.edu.sg).}}
\author{\IEEEauthorblockN{Qin~Tao, Shuowen~Zhang,~\IEEEmembership{Member,~IEEE},  Caijun~Zhong,~\IEEEmembership{Senior Member,~IEEE}, and Rui~Zhang,~\IEEEmembership{Fellow,~IEEE}}}
\maketitle
\vspace{-3mm}
\begin{abstract}
In this letter, we consider a multicast system where a single-antenna transmitter sends a common message to multiple single-antenna users, aided by an intelligent reflecting surface (IRS) equipped with $N$ passive reflecting elements. Prior works  on IRS have mostly assumed the availability of channel state information (CSI) for designing its passive beamforming. However, the acquisition of CSI requires substantial training overhead that increases with $N$. In contrast, we propose in this letter a novel \emph{random passive beamforming} scheme, where the IRS performs independent random reflection for $Q\geq 1$ times in each channel coherence interval without the need of CSI acquisition. For the proposed scheme, we first derive a closed-form approximation of the outage probability, based on which the optimal $Q$ with best outage performance can be efficiently obtained. Then, for the purpose of comparison, we derive a lower bound of the outage probability with traditional CSI-based passive beamforming. Numerical results show that a small $Q$ is preferred in the high-outage regime (or with high rate target) and the optimal $Q$ becomes larger as the outage probability decreases (or as the rate target decreases). Moreover, the proposed scheme significantly outperforms the CSI-based passive beamforming scheme with training overhead taken into consideration when $N$ and/or the number of users are large, thus offering a promising CSI-free alternative to existing CSI-based schemes.
\end{abstract}
\vspace{-1mm}
\keywords  Intelligent reflecting surface, multicast, random passive beamforming, outage probability

\vspace{-5mm}
\section{Introduction}
\vspace{-1mm}
Intelligent reflecting surface (IRS) is a new emerging paradigm for the fifth-generation (5G) and beyond wireless communication networks \cite{WuMag,BasarSurvey,tutorial}. Specifically, an IRS is able to alter the wireless channel by shifting the phases of the impinging signals via a large number of \emph{passive} reflecting elements, for enhancing desired signal power or suppressing undesired interference. Being also cost-effective and energy-efficient, IRS has received fast-growing research attention recently. To maximize the performance gains brought by IRS, it is of paramount importance to properly design the IRS reflection coefficients, which is also termed as \emph{passive beamforming}. In the literature, passive beamforming design has been studied under various system setups with different performance metrics \cite{WuTWCjoint,shuowen,XH,OFDM,OFDMA,CPan1,ZHOU,SPAWC,twoscale}. Particularly, the existing works on IRS have mostly considered a ``fully intelligent'' reflecting surface with the availability of full channel state information (CSI) for all the IRS-associated links (see, e.g., \cite{WuTWCjoint,shuowen,XH,OFDM,OFDMA,CPan1,ZHOU,SPAWC}). In practice, such CSI needs to be acquired in each channel coherence interval by the transmitter/receiver at the cost of channel training overhead and/or feedback complexity that increases with the number of IRS reflecting elements, $N$, which is prohibitive since $N$ is typically very large for IRS \cite{OFDM}. This inevitably results in delay and reduced throughput. On the other hand, \cite{twoscale} has proposed a two-timescale passive beamforming design based on the statistical CSI, which, however, still requires estimating the channel statistics for which the complexity increases with $N$.

\begin{figure}[t]
	\centering
	\includegraphics[width=8cm]{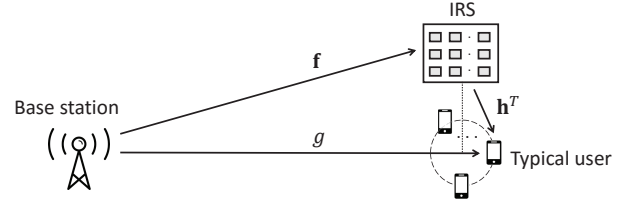}
	\vspace{-3mm}
	\caption{An IRS-aided multi-user multicast system.}\label{fig:systemmodel_1}
	\vspace{-1mm}
\end{figure}

To alleviate the CSI acquisition burden, we propose in this letter a novel ``\emph{random passive beamforming}'' scheme, where the IRS only performs random reflection without the need of knowing any CSI. Specifically, we consider an IRS-aided multi-user multicast system, as illustrated in Fig. \ref{fig:systemmodel_1}, where a single-antenna base station (BS) aims to send a common message to multiple single-antenna users within a local cluster, aided by an IRS with $N$ reflecting elements. In practice, the IRS is usually deployed in the vicinity of either the BS or the users to minimize its double path loss with them. In this letter, we assume that the IRS is deployed in the vicinity of the user cluster so as to smartly adjust itself to serve this cluster without affecting other clusters. Each channel coherence interval is divided into $Q\geq 1$ ``reflecting slots'', and an independent random passive beamforming pattern is applied in each slot. Note that different from the traditional (active) random beamforming for broadcast systems where the multi-antenna BS performs random beamforming to exploit the multi-user channel diversity via CSI-based transmission scheduling \cite{RB}, we aim to smartly reshape the \emph{distribution} of all users' channels aided by the IRS in a multicast system when CSI is unavailable. Moreover, compared to the CSI-based passive beamforming, the proposed scheme only requires knowledge of the effective BS-user channel in each slot, hence the training overhead scales with $Q$ rather than $N$, which is greatly reduced with $Q\!\ll\! N$. However, it is yet unclear how fast the channel should change to achieve the most desirable distribution (i.e., finding the optimal value of $Q$), and how the proposed random passive beamforming performs as compared to the designed beamforming in the ideal case with perfect CSI when considering the channel training overhead. To answer these questions, we first analyze the \emph{outage probability} of each user under the proposed scheme. In particular, we derive an accurate approximation for the outage probability in closed-form, and draw useful insights into the selection of the optimal $Q$. For comparison, we also provide a lower bound of the outage probability with the CSI-based scheme. Numerical results show that a small $Q$ is optimal when the outage probability is high (or the rate target is high), and the optimal $Q$ increases as the outage probability decreases (or as the rate target decreases). Moreover, the proposed scheme achieves significant performance gains over the CSI-based scheme with training overhead when $N$ and/or the number of users in the system are large.

\begin{figure}[t]
	\centering
	\includegraphics[width=7cm]{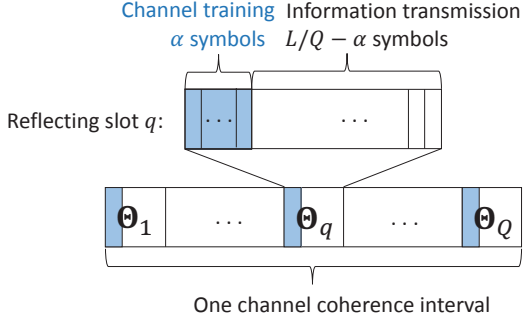}
	\vspace{-3mm}
	\caption{The proposed random passive beamforming scheme.}\label{fig:systemmodel_2}
	\vspace{-1mm}
\end{figure}

\vspace{-2mm}
\section{System Model}
\vspace{-1mm}
We consider an IRS-aided multicast system shown in Fig. \ref{fig:systemmodel_1}, where a single-antenna BS aims to send a common message to $K\geq 1$ single-antenna users that are located within a cluster, and an IRS consisting of $N$ reflecting elements is deployed in the vicinity of the user cluster to reflect the signals from the BS to all users for enhancing the information multicasting performance. Different from the existing works on CSI-based passive beamforming, we propose a \emph{random passive beamforming} scheme with no CSI needed, where each channel coherence interval is equally divided into $Q\geq 1$ ``reflecting slots'', where $Q\in \mathcal{Q}=\{1,...,Q_{\max}\}$ with $Q_{\max}$ being the maximum number of times for IRS reflection pattern change; and the IRS reflects with a \emph{random} set of coefficients over each time slot, as illustrated in Fig. \ref{fig:systemmodel_2}. Let ${\bf {\Theta}}_q\!=\!{\text{diag}}\{e^{j\theta_{1,q}}, e^{j\theta_{2,q}},...,e^{j\theta_{N,q}}\}$ denote the IRS reflection matrix at each $q$th reflecting slot, $q\!=\!1,...,Q$, where $\theta_{n,q} \in [0,2\pi)$ represents the phase shift at the $n$th reflecting element and is generated based on an independent {{random variable (RV)}} uniformly distributed in $[0,2\pi)$. {{Let $g_k\in \mathbb{C}$ denote the direct channel from the BS to user $k$, ${\bf{h}}_k^T\in \mathbb{C}^{1\times N}$ denote the channel vector from the IRS to user $k$, and ${\bf{f}}\in \mathbb{C}^{N\times 1}$ denote the channel vector from the BS to the IRS. We assume that ${\bf{f}}$ follows the line-of-sight (LoS) model specified in \cite{shuowen} by properly deploying the IRS, where the channel from the BS to each IRS element has different phase but same amplitude $\sigma_f$ (i.e., $|f_1|=...=|f_N|=\sigma_f$) since the BS-IRS distance is generally much larger than the IRS size. For the purpose of exposition, we present the remaining channel models and achievable rate for one typical user indexed by $\bar{k}$ as follows, whose corresponding direct BS-user channel and IRS-user channel are denoted by $g=g_{\bar{k}}$ and ${\bf{h}}^T={\bf{h}}_{\bar{k}}^T$, respectively.}} Specifically, we consider a quasi-static Rayleigh fading channel model for $g$ and ${\bf{h}}^T$, where they remain approximately static in each coherence interval, but may change independently among different coherence intervals; while in each coherence interval, $g$ and each element in ${\bf{h}}^T$ are modeled as an independent circularly symmetric complex Gaussian (CSCG) RV with average power $\sigma_g^2$ and $\sigma_h^2$, respectively, i.e., $g\sim\mathcal{CN}(0,\sigma_g^2)$ and ${\bf{h}}\sim\mathcal{CN}({\bf {0}},\sigma_h^2{\bf I}_N)$.

At each $q$th reflecting slot, the received signal at the user is the sum of the transmitted signal via the direct channel and that via the IRS-reflected channel, which is given by
\vspace{-2mm}\begin{equation}
y=\sqrt{P}(g+{\bf{h}}^T{\bf{\Theta}}_{q}{\bf{f}})s+z,\quad q\!=\!1,...,Q,
\vspace{-2mm}\end{equation}
where $P$ denotes the average transmission power at the BS; $s\sim \mathcal{CN}(0,1)$ denotes the transmitted symbol modeled by a CSCG random variable with zero mean and unit variance; and $z\sim \mathcal{CN}(0,\sigma_z^2)$ denotes the CSCG noise at the user receiver with average power $\sigma_z^2$. { {We assume that all users are able to perfectly estimate the effective channel at each reflecting slot (e.g., $g+{\bf{h}}^T{\bf{\Theta}}_{q}\bf{f}$ for the typical user) based on $\alpha\geq1$ training symbols sent by the BS.}}\footnote{Note that all users have approximately the same received SNR level since they are located closely within a cluster, thus requiring roughly the same number of training symbols to achieve a given channel estimation accuracy.} However, it is worth noting that each user does not need knowledge of the individual channels (e.g., $g$, ${\bf{h}}^T$, and ${\bf{f}}$ for the typical user) in each channel coherence interval, the acquisition of which requires training overhead that scales with $N$ \cite{OFDM}. Let $L$ denote the total number of symbols in each coherence interval.\footnote{We assume that $L/Q$ is an integer and $L\!\gg\! \alpha Q$ in practice.} Thus, the fraction of time for information transmission at each reflecting slot over the entire channel coherence interval is $\frac{1}{Q}(1-\frac{\alpha}{L/Q})=\frac{1}{Q}-\frac{\alpha}{L}$. Therefore, the maximum achievable rate in each coherence interval for the typical user can be expressed as
\vspace{-1mm}\begin{equation}\label{rate}
R=\bigg(\frac{1}{Q}-\frac{\alpha}{L}\bigg)\sum_{q=1}^Q\log_2 \bigg(1+\gamma|g+{\bf{h}}^T{\bf {\Theta}}_{q}{\bf{f}}|^2\bigg),
\vspace{-1mm}\end{equation}
in bits per second per Hertz (bps/Hz), where $\gamma\!=\!P/\sigma_z^2$ denotes the reference SNR.

In this letter, we consider delay-limited communication and thus focus on the analysis of \emph{outage probability} to measure the reliability of the typical user. The outage probability for a given rate target $\tau$ (in bps/Hz) is defined as
\vspace{-2mm}
\begin{equation}
P_{\text{out}}\!=\!\mathbb{P} \bigg\{\bigg(\frac{1}{Q}\!-\!\frac{\alpha}{L}\bigg)\sum_{q=1}^{Q}\log_2 \left(1\!+\!\gamma|g+{\bf{h}}^T{\bf {\Theta}}_{q}{\bf{f}}|^2\right)\!<\!\tau\bigg\}. \label{POUTK}
\vspace{-2mm}
\end{equation}
Note that there exists an interesting trade-off in $P_{\text{out}}$ by tuning the number of reflecting slots, $Q$: as $Q$ increases, more variation (fading) is introduced to the effective channel gain, thus providing more chance to avoid severe outage since the channels in deep fade are more likely to be averaged out; on the other hand, more training time (i.e., $\alpha Q$ symbols) is needed for estimating the channels over reflecting slots. Therefore, it is generally unclear how we should set $Q$ to minimize the outage probability. In the rest of this letter, we analyze the outage probability in (\ref{POUTK}) to draw insights into the optimal choice of $Q$; moreover, we compare the outage performance of the proposed scheme with the traditional CSI-based passive beamforming scheme.

\vspace{-2mm}
\section{Outage Probability Analysis of Random Passive Beamforming}\label{transmission}
In this section, we analyze the outage probability of the proposed random passive beamforming scheme. Note that $P_{\mathrm{out}}$ in (\ref{POUTK}) is determined by the joint distribution of the channel coefficients in $g$ and ${\bf{h}}^T$, as well as the random reflection coefficients, $\{{\bf{\Theta}}_q\}$, which are coupled in a complicated manner. Thus, it is generally difficult to obtain the exact closed-form expression for $P_{\mathrm{out}}$. Motivated by this and the practically high signal-to-noise ratio (SNR) regimes that information multicasting service typically operates in (for meeting the high data rate requirements such as $4$K video streaming), we focus on the high-SNR regime with large $\gamma$ and characterize the corresponding asymptotic expression for $P_{\mathrm{out}}$ in the following.

Specifically, when $\gamma$ is large, the element ``$1$'' inside the logarithm of (\ref{POUTK}) can be omitted, thus ${P}_{\text{out}}$ can be well-approximated as
\vspace{-2mm}\begin{equation}
\!P_{\mathrm{out}}\!\approx\!\mathbb{P}\bigg\{\bigg(\!\frac{1}{Q}-\frac{\alpha}{L}\!\bigg)\sum_{q=1}^{Q}\log_2 \!\bigg(\gamma|g+{\bf{h}}^T{\bf {\Theta}}_{q}{\bf{f}}|^2\bigg)\!\!<\!\!\tau\bigg\}\!\overset{\Delta}{=}\!\tilde{P}_{\text{out}}.\!\!\!\!
\vspace{-2mm}\end{equation}
Moreover, $\tilde{P}_{\text{out}}$ can be further simplified as
\vspace{-2mm}\begin{equation}\label{POUTLO}
\tilde{{P}}_{\text{out}}\!=\!\mathbb{P}\bigg\{\prod_{q=1}^{Q} \!|g\!+\!{\bf{h}}^T{\bf {\Theta}}_{q}{\bf{f}}|^2\!<\!2^{{\tau}/{\big(\frac{1}{Q}-\frac{\alpha}{L}\big)}}/{\gamma}^Q\bigg\}.
\vspace{-2mm}\end{equation}
Notice from (\ref{POUTLO}) that $\tilde{{P}}_{\text{out}}$ is critically dependent on $V\overset{\Delta}{=}\prod_{q=1}^{Q} |g+{\bf{h}}^T{\bf {\Theta}}_{q}{\bf{f}}|^2$, which is the product of $Q$ correlated exponential RVs. Thus, the exact distribution of $V$ is generally difficult to obtain. Nevertheless, we show below that such correlation is negligible under certain practical conditions.

\begin{lemma}\label{lem1}
With $N\gg\sigma_g^2/(\sigma_h^2\sigma_f^2)$, the channel gains at different reflecting slots, $\!|g\!+\!{\bf{h}}^T{\bf {\Theta}}_{q}{\bf{f}}|^2$, $q\!=\!1,...,Q$, can be approximated as \emph{independent} exponential RVs.
\end{lemma}
\proof
First, it can be easily shown that $X_q\!\overset{\Delta}{=}\!g+{\bf{h}}^T{\bf {\Theta}}_{q}{\bf{f}}$'s are identically distributed Gaussian RVs with common mean $\mathbb{E}[X_q]=0$ and variance $\mathrm{Var}[X_q]=\sigma_g^2+N\sigma_h^2\sigma_f^2$, $\forall q$. The Pearson product-moment correlation coefficient between any two $X_{q_1}$ and $X_{q_2}$ with $q_1\neq q_2$ can be thus obtained as
\begin{align} \label{PMCC}
\!\!\!&\rho _{q_1,q_2} \!=\!\mathbb{E}[(X_{q_1}\!-\!\mathbb{E}[X_{q_1}])(X_{q_2}\!-\!\mathbb{E}[X_{q_2}])^*]/\sqrt{\mathrm{Var}[X_{q_1}]\mathrm{Var}[X_{q_2}]}\nonumber\\
&\!=\! \frac{\sigma_g^2\!+\!\sigma_h^2\sigma_f^2\sum_{n=1}^{N}\mathbb{E}[e^{j\theta_{n,q_1}}]\mathbb{E}[e^{-j\theta_{n,q_2}}]}{\sigma_g^2+N\sigma_h^2\sigma_f^2}\!=\!\frac{\sigma_g^2}{\sigma_g^2\!+\!N\sigma_h^2\sigma_f^2},\!\!\!
\end{align}
where $\mathbb{E}[e^{j\theta_{n,q}}]=0,\ \forall n,\forall q$ holds since $\theta_{n,q}$ is uniformly distributed in $[0,2\pi)$. Based on (\ref{PMCC}), when $N\gg \sigma_g^2/(\sigma_h^2\sigma_f^2)$ or equivalently $\sigma_g^2\ll N\sigma_h^2\sigma_f^2$ holds, i.e., the reflected channel power is much larger than the direct channel power, we have $\rho_{q_1,q_2}\approx 0$. This thus indicates that $X_q$'s are independent Gaussian RVs, and consequently $|X_q|^2$'s are independent exponential RVs.
\endproof
 {Note that the intuition behind Lemma \ref{lem1} lies in the fact that the reflected channels ${\bf{h}}^T{\bf {\Theta}}_{q}{\bf{f}}$'s at different reflecting slots are independent of each other regardless of the values of $N$ and $Q$, due to the independent passive beamformers ${\bf {\Theta}}_{q}$'s. Thus, when $N$ is sufficiently large such that the reflected channel becomes the dominant path, the overall channels $X_q$'s will also be approximately independent.} By leveraging the results in Lemma \ref{lem1}, a closed-form approximation of $\tilde{P}_{\text{out}}$ can be obtained as follows.

\begin{figure}[t]
	\centering
	\includegraphics[width=6cm]{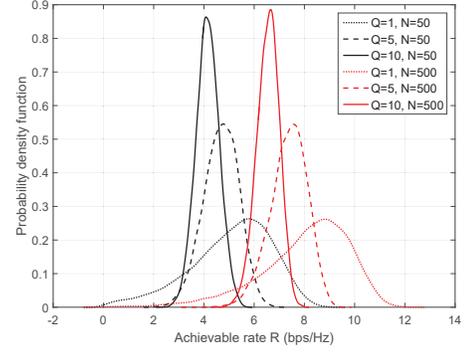}
	\vspace{-3mm}
	\caption{PDF of the achievable rate $R$.}\label{fig:C_distribution}
\end{figure}

\begin{theorem}\label{lemma3}
With $N\gg\sigma_g^2/(\sigma_h^2\sigma_f^2)$, $\tilde{P}_{\text{out}}$ can be well approximated as
\vspace{-2mm}\begin{equation}\label{CDFprod}
\tilde{{P}}_{\text{out}}\!\approx\! G
{\setlength{\arraycolsep}{0.1pt}
\scriptsize
\begin{array}{ccc}
 Q&1\\
1& Q+1
\end{array}
}
\bigg({2^{{\tau}/{\big(\frac{1}{Q}-\frac{\alpha}{L}\big)}}}/{(\gamma\lambda)^Q}\bigg|
\begin{array}{ccc}
 1\\
{{\bf{1}}_{Q}^T,0}
\end{array}\bigg),\!
\vspace{-2mm}\end{equation}
where $\lambda\!=\!\sigma_g^2\!+\!N\sigma_h^2\sigma_f^2$; $G
{\setlength{\arraycolsep}{0.1pt}
\scriptsize
\begin{array}{ccc}
 m&n\\
p&q
\end{array}
}
\bigg(x\bigg|
\begin{array}{ccc}
 a_1,...,a_p\\
b_1,...b_q
\end{array}\bigg)$ represents the Meijer G-function \cite[9.301]{Table}; and ${\bf{1}}_{Q}^T$ is an all-one row vector with length $Q$.
\end{theorem}
\proof
According to Lemma \ref{lem1}, with $N\gg\sigma_g^2/(\sigma_h^2\sigma_f^2)$, $V=\prod_{q=1}^Q |X_q|^2$ can be approximately regarded as the product of $Q$ independent and identically distributed (i.i.d.) exponential RVs each with mean $\lambda$, whose probability density function (PDF) is characterized as
$f_{V}(v)\!\approx\! G
\setlength{\arraycolsep}{0.1pt}
\scriptsize
\begin{array}{ccc}
Q&0\\
0&Q
\end{array}
\bigg(\frac{x}{\lambda^{Q}}\bigg|0\bigg)$ \cite{productpdf}. Based on this, $\tilde{{P}}_{\text{out}}$ in (\ref{POUTLO}) can be expressed as $\tilde{{P}}_{\text{out}}\!=\! \int_0^{2^{{\tau}/{(\frac{1}{Q}-\frac{\alpha}{L})}}/{\gamma}^Q} f_{V}(v)\text{d}v$. The results in Theorem \ref{lemma3} thus follows by applying the integration relationship in \cite[Eq. 7.811.2]{Table}.
\endproof
The expression of $\tilde{P}_{\text{out}}$ in Theorem \ref{lemma3} is complicated since it involves the Meijer G-function, based on which the impact of $Q$ is difficult to analyze. Nevertheless, the closed-form nature of $\tilde{P}_{\text{out}}$ enables efficient search for the optimal $Q$, since $Q$ is an integer with a finite feasible set $\mathcal{Q}$. Specifically, based on the system parameters (i.e., $P$, $N$, $L$, $\alpha$, $\tau$ and $\lambda$), {\footnote{{Note that these parameters can be obtained \emph{a priori} at the BS, e.g., through control/feedback links from the IRS (e.g., $N$) or users (e.g., $\lambda$).}}} the BS can compute the approximate outage probability $\tilde{P}_{\text{out}}$ corresponding to every feasible value of $Q\in \mathcal{Q}$ through (\ref{CDFprod}), and select the optimal $Q$ denoted by $\tilde{Q}^\star$ with the lowest $\tilde{P}_{\text{out}}$. The value of $\tilde{Q}^\star$ is then sent to the IRS for performing random passive beamforming. In Section \ref{sec_num}, we will numerically verify the accuracy of the approximation of the exact outage probability ${P}_{\text{out}}$ with $\tilde{P}_{\text{out}}$, as well as that of the optimal $Q$ that minimizes ${P}_{\text{out}}$ with $\tilde{Q}^\star$.

 {Furthermore, to obtain more intuitive insights into the impact of $Q$ on the exact outage probability $P_{\mathrm{out}}$, we show in Fig. \ref{fig:C_distribution} the PDF of the exact achievable rate $R$ with different values of $Q$.\footnote{We set $P=20$ dBm for this numerical example, and other parameters will be given later in Section \ref{sec_num}.} It can be observed that as $Q$ increases, both the mean and variance of $R$ are reduced. As such, to minimize the outage probability $P_{\mathrm{out}}$, a small $Q$ is desirable when the rate target is high (corresponding to the high-outage or low-SNR regime), while a large $Q$ is desirable when the rate target is low (corresponding to the low-outage or high-SNR regime).} In addition, we also show in Fig. \ref{fig:C_distribution} the PDF of $R$ with different numbers of IRS reflecting elements, $N$. It is observed that as $N$ increases, the PDF of $R$ moves right. This suggests that the outage probability $P_{\mathrm{out}}$ generally decreases with $N$, due to stronger effective channel.

\section{CSI-based Multicast Beamforming}
In this section, we introduce a CSI-based passive beamforming scheme for comparison with our proposed scheme. Note that the IRS reflection coefficient design only depends on the direct channel and cascaded IRS-reflected channels for the $K$ users (i.e., $f_kh_{kn}$'s). Thus, we assume that the BS sends a training sequence to all $K$ users, based on which each user estimates its direct channel and cascaded reflected channels (i.e., $g_k$ and $f_kh_{kn}$'s) and feeds them back to the BS. The required number of training symbols for estimating these $N+1$ complex channel coefficients generally scales with $N$ \cite{OFDM}, and for fair comparison we assume that $\alpha(N+1)$ symbols (with the same training overhead $\alpha$ as the proposed random passive beamforming scheme) suffice for perfect CSI estimation; while the feedback time overhead/delay is omitted for simplicity.\footnote{On the other hand, the CSI can also be obtained through reverse-link channel training by letting the users send training sequences to the BS without the need of feedback. However, the required training time generally scales with $K(N+1)$, which will lead to worse outage performance compared to the forward-link channel training considered above.} Therefore, the maximum fraction of time for information transmission over each coherence interval is $\big(1-\frac{\alpha(N+1)}{L}\big)$.

Based on the obtained CSI, the BS designs the passive beamformer $\bf{\Theta}$ to minimize the maximum outage probability among the $K$ users. For any given rate target $\tau$, the outage probability at the $k$th user is given by\footnote{Note that the channel training overhead can be reduced by grouping $M$ adjacent IRS elements as one effective element \cite{OFDM}. With $\bar{N}=N/M$ denoting the total number of effective elements, the corresponding outage probability is given by (\ref{PoutCSI}) with $N$ substituted by $\bar{N}$ and $\bf{\Theta}$ by an $\bar{N}\times \bar{N}$ diagonal matrix.}
\begin{align}\label{PoutCSI}
\!\!\!\!\!P_{\text{out},k}\!=\! \mathbb{P}\left\{\!\left(\!1\!-\!\frac{\alpha(N\!+\!1)}{L}\!\right)\log_2\left(1\!+\!\gamma|g_k\!+\!{\bf{h}}_k^T{\bf {\Theta}}{\bf{f}}|^2\right)\!<\!\tau\!\right\}\!.\!
\end{align}
Thus, the optimization problem is formulated as
\begin{equation}
\mbox{(P1)} \  \max_{{\bf{\Theta}}:|[{\bf{\Theta}}]_{n,n}|=1,\forall n}\ \min_k\ |g_k+{\bf{h}}_k^T{\bf {\Theta}}{\bf{f}}|^2.
\end{equation}
Note that (P1) is similar to the constant envelope precoding optimization problem for multicasting systems \cite{CEmulticast}, thus the similar semi-definite relaxation (SDR) based algorithm proposed in \cite{CEmulticast} can be applied for finding both an upper bound of the optimal value of (P1) and a suboptimal solution to (P1), which corresponds to a lower and upper bound of the worst-case outage probability of the CSI-based scheme, respectively. Due to limited space, we omit the details for brevity.

Note that since $N$ is practically very large, the training overhead of the CSI-based scheme is generally much larger compared to our proposed scheme; while on the other hand, the CSI-based passive beamforming design is anticipated to outperform the random passive beamforming due to the exploitation of CSI. Hence, whether or not this benchmark scheme will outperform the proposed scheme remains unknown. To answer this question, we compare the outage performance of the CSI-based scheme with our proposed scheme by simulation in the next section.

\section{Numerical results}\label{sec_num}
In this section, we provide numerical results to validate our analysis. The distance-dependent path loss for each link is modeled as $\sigma^2_{i}\!=\!\sigma_0^2({d_i}/{d_0})^{-\alpha_i}$, $i\in\{{g,{{f}},{{h}}}\}$, where $\sigma_0^2\!=\!-30$ dB denotes the reference path loss at $d_0\!=\!1$ m, $d_i$ and $\alpha_i$ denote the distance and path loss exponent of each link. We assume that the BS and IRS are located on a three-dimensional plane with coordinates $[0,0,2]$ m and $[100,0,2]$ m, respectively. The typical user is located at height $1$ m right below the IRS. Thus, we have $d_{{{f}}}\!=\!100$ m, $d_{g}\!\approx\!100$ m, and $d_{{{h}}}\!=\!1$ m. We further set $\alpha_g\!=\!3.5$, $\alpha_{{{f}}}\!=\!2$, and $\alpha_{{{h}}}\!=\!2.5$. The average noise power is set as $\sigma_z^2\!=\!-90$ dBm. In addition, we set the total number of symbols in each coherence interval as $L\!=\!1000$ and the number of training symbols used for estimating one channel coefficient as $\alpha\!=\!20$. The maximum value of $Q$ is set as $Q_{\max}=8$.

\begin{figure}[t]
	\centering
	\includegraphics[width=7cm]{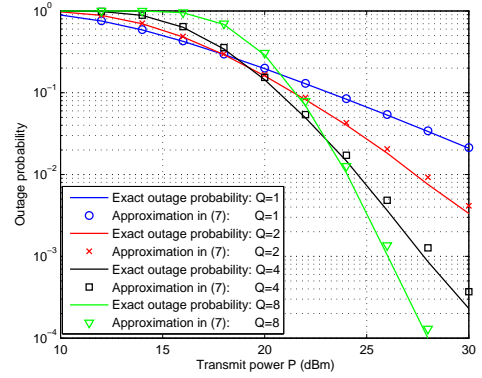}
	\vspace{-3mm}
	\caption{ {Outage probability versus transmit power.}}\label{fig:fig8b}
	\vspace{-3mm}
\end{figure}

First, we focus on the typical user and plot in Fig. \ref{fig:fig8b} both the accurate outage probability $P_{\mathrm{out}}$ and its approximation $\tilde{P}_{\mathrm{out}}$ versus the transmit power $P$ with  {rate target $\tau\!=\!6$ bps/Hz and $N=300$}. It can be observed that the approximation  of $P_{\mathrm{out}}$ with $\tilde{P}_{\mathrm{out}}$ is already tight for all values of $P$ and $Q$ when $N=300$, which indicates that the approximation proposed in Theorem \ref{lemma3} is accurate. Moreover, it can be observed that the optimal value of $Q$ varies with the value of $P$.

\begin{figure}[t]
	\centering
	\includegraphics[width=6cm]{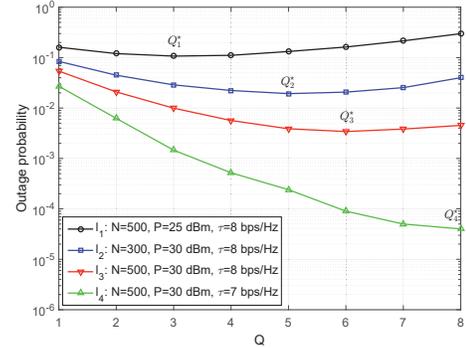}
	\vspace{-3mm}
	\caption{Outage probability versus $Q$.}\label{fig:optKtradeoff}
	\vspace{-1mm}
\end{figure}

\begin{figure}[t]
	\centering
	\includegraphics[width=6cm]{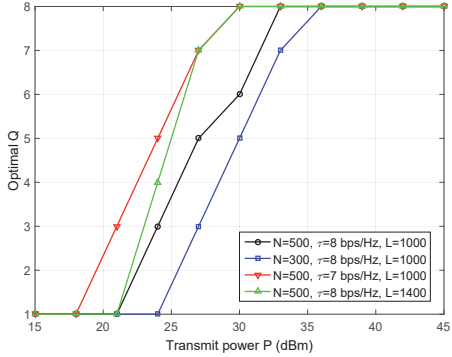}
	\vspace{-3mm}
	\caption{Optimal $Q$ versus transmit power.}\label{fig:optK}
\end{figure}

To further investigate the optimal $Q$ with different system parameters, we show in Fig. \ref{fig:optKtradeoff} the outage probability versus $Q$ with different values of $N$, $P$, and $\tau$. The optimal $Q$ for curves $l_i$, $i=1,2,3,4$, is denoted as $Q_i^\star$. It can be clearly observed that the outage probability generally first decreases and then increases as $Q$ becomes larger, thus validating the trade-off between increased training overhead and enhanced channel diversity in the overall outage performance. In addition, it is observed that the optimal $Q$ increases with $N$ and transmit power $P$, but decreases with the rate target $\tau$. In Fig. \ref{fig:optK}, we further show the optimal $Q$ versus $P$ with different values of $N$, $\tau$, and $L$. It can be observed that the outage probability is minimized by setting $Q\!=\!1$ in the low-SNR regime (with high outage probability); while in the moderate-SNR regime, the optimal $Q$ increases with the SNR, until it reaches the maximum value $Q_{\max}\!=\!8$, which are consistent with our analysis in Section III. In addition, it is more clearly observed that the optimal $Q$ increases with $N$, but decreases with the number of symbols in each coherence interval, $L$, and the target rate $\tau$. The above results indicate that the IRS should properly adjust the times of random reflection, $Q$, in practice.

\begin{figure}[t]
	\centering
	\includegraphics[width=7cm]{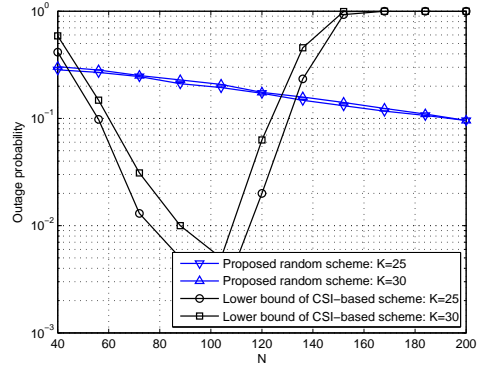}
	\vspace{-3mm}
	\caption{ {Comparison of the proposed scheme with the benchmark scheme.}}\label{fig:versus}
	\vspace{-1mm}
\end{figure}

Finally, in Fig. \ref{fig:versus}, we compare the the maximum (worst-case) outage probability in the system with our proposed scheme versus the lower bound of that with the CSI-based passive beamforming scheme proposed in Section IV.  {To reduce the training overhead of the CSI-based scheme, we group $M=8$ IRS elements as an effective element and perform corresponding beamforming design \cite{OFDM}. In addition, we set $P\!=\!31$ dBm, $Q\!=\!4$, $\tau\!=\!5$ bps/Hz, and $K$ users randomly distribute on a circle with radius $5$ m centered at the IRS.}  {It is observed that increasing the number of users has little effect on the outage performance of the proposed scheme due to the similar BS-user distances, but significantly deteriorates that of the CSI-based scheme since the efficacy of passive beamforming degrades when more users need to be catered for. Moreover, it is observed that the outage probability of the proposed random passive beamforming scheme decreases monotonically as $N$ increases; while the outage probability for the CSI-based scheme first decreases and then increases with $N$, since a larger $N$ results in enhanced beamforming gain but increased training overhead as well. Particularly, when $N$ becomes large, the performance of the CSI-based scheme is dominated by the training overhead, and our proposed scheme achieves significantly improved performance. The above results indicate that the proposed random passive beamforming is particularly favorable for the practical scenario with large number of reflecting elements, $N$, and/or large number of users, $K$.}

\section{Conclusions}
In this letter, we proposed a novel random passive beamforming scheme for an IRS-aided multi-user multicast system, where the IRS reflection pattern is randomly changed for $Q$ times in each channel coherence interval without the need of CSI acquisition. For the proposed scheme, we derived a closed-form approximate expression for its outage probability in the high-SNR regime, based on which the optimal $Q$ can be efficiently obtained. Numerical results validated the accuracy of our approximation of the outage probability. Moreover, it was revealed that a small $Q$ is suitable in the high-outage regime, while a larger $Q$ is preferred as the outage probability decreases. Furthermore, the proposed scheme also outperforms the traditional CSI-based passive beamforming scheme, especially when the number of IRS reflecting elements and/or users are large, which validates the practical usefulness of the proposed random passive beamforming scheme.

\bibliographystyle{IEEE}
\begin{spacing}{0.9}

\end{spacing}
\end{document}